\documentstyle[11pt,amssymb]{article}

\def\dalemb#1#2{{\vbox{\hrule height .#2pt
        \hbox{\vrule width.#2pt height#1pt \kern#1pt
                \vrule width.#2pt}
        \hrule height.#2pt}}}

\def\0{{\sst{(0)}}}
\def\1{{\sst{(1)}}}
\def\2{{\sst{(2)}}}
\def\3{{\sst{(3)}}}
\def\4{{\sst{(4)}}}
\def\5{{\sst{(5)}}}
\def\6{{\sst{(6)}}}
\def\7{{\sst{(7)}}}
\def\8{{\sst{(8)}}}

\def\wtd{\widetilde}

\def\nn{\nonumber} \def\bd{\begin{document}} \def\ed{\end{document}}
\def\ds{\documentstyle} \let\fr=\frac \let\bl=\bigl \let\br=\bigr
\let\Br=\Bigr \let\Bl=\Bigl
\let\bm=\bibitem
\let\na=\nabla
\let\pa=\partial \let\ov=\overline
\newcommand{\be}{\begin{equation}}
\newcommand{\ee}{\end{equation}}
\def\ba{\begin{array}}
\def\ea{\end{array}}
\def\ft#1#2{{\textstyle{{\scriptstyle #1}\over {\scriptstyle #2}}}}
\def\fft#1#2{{#1 \over #2}}
\def\del{\partial}
\def\sst#1{{\scriptscriptstyle #1}}
\def\oneone{\rlap 1\mkern4mu{\rm l}}
\def\ie{{\it i.e.\ }}
\def\etc{{\it etc.\ }}
\def\via{{\it via}}
\def\semi{{\ltimes}}
\def\cv{{\cal V}}
\def\str{{\rm str}}
\def\jm{{\rm j}}
\def\im{{\rm i}}

\def\cramp{\medmuskip = 2mu plus 1mu minus 2mu}
\def\cramper{\medmuskip = 2mu plus 1mu minus 2mu}
\def\crampest{\medmuskip = 1mu plus 1mu minus 1mu}
\def\uncramp{\medmuskip = 4mu plus 2mu minus 4mu}

\def\mapright#1{\smash{\mathop{-\!\!\!-\!\!\!-\!\!\!-\!\!\!-\!\!\!
             \longrightarrow}\limits^{#1}}}
\def\maprightt#1#2{\smash{\mathop{-\!\!\!-\!\!\!-\!\!\!-\!\!\!-\!\!\!
             \longrightarrow}\limits^{#1}_{#2}}}

\def\tX{{{\wtd X}}}
\newcommand{\ho}[1]{$\, ^{#1}$}
\newcommand{\hoch}[1]{$\, ^{#1}$}
\newcommand{\bea}{\begin{eqnarray}}
\newcommand{\eea}{\end{eqnarray}}
\newcommand{\ra}{\rightarrow}
\newcommand{\lra}{\longrightarrow}
\newcommand{\Lra}{\Leftrightarrow}
\newcommand{\ap}{\alpha^\prime}
\newcommand{\bp}{\tilde \beta^\prime}
\newcommand{\tr}{{\rm tr} }
\newcommand{\Tr}{{\rm Tr} }
\newcommand{\NP}{Nucl. Phys. }
\newcommand{\tamphys}{\it Center for Theoretical Physics\\
Texas A\&M University, College Station, Texas 77843}
\newcommand{\upenn}{\it Department of Physics and Astronomy\\
University of Pennsylvania, Philadelphia, Pennsylvania 19104}

\newcommand{\auth}{M. Cveti\v{c}\footnote{Based on the talk given at
Strings'2000.}
 }

\begin{document}
\begin{flushright}
\hfill{UPR-915-T}\\
\hfill{hep-th/0012105}\\
\hfill{December 2000}\\
\end{flushright}

\vspace{15pt}

\begin{center}{ \large {\bf Domain Wall World(s)
 }}

\vspace{15pt}
\auth

\vspace{15pt}

{\hoch{\dagger}\upenn}

\

\vspace{40pt}

\underline{ABSTRACT}
\end{center}

Gravitational properties of domain walls  in fundamental theory and their
implications for the trapping of gravity are reviewed.
In particular,   the difficulties  to embed gravity trapping configurations
within gauged supergravity is reviewed and 
the status of the domain walls obtained via  the breathing mode
 of  sphere reduced Type IIB supergravity is
presented.

{\vfill\leftline{}\vfill
\vskip 5pt

\pagebreak \setcounter{page}{1}  Over the past few years domain
walls have played an  important role both from the point of view
of the AdS/CFT correspondence, i.e. by shedding light on
the renormalization group flow  and bound state
spectra of strongly
coupled field theories,  as well as  from the point of view of the
brane world scenarios, such as Randall-Sundrum scenario  \cite{rs2} for
localization of gravity on domain walls in five-dimensions.
In this contribution, however,  we shall focus on  gravitational properties
of  domain walls, their implications for  the brane world
scenarios  and  their implementation in the
fundamental theory.

This presentation is based on the work with K. Behrndt
\cite{bc1,bc2}, along with the parallel work by Kallosh and Linde
and Schmakova \cite{kls,kl},  leading to the
``no-go'' theorems
for the implementation of the Randall-Sundrum scenario, when one  
employs 
massless modes of gauged supergravity. (For  subsequent 
developments see, e.g.,  \cite{gz,cd,bhl}.)
The focal part,  however, will be  based on the work with H. L\" u
and C. Pope  \cite{clp1,clp2} where the 
massive-breathing mode of sphere reduced gauged supergravity 
 was proposed
 as the candidate field  for the  gravity trapping domain walls.
For  subsequent related  works see \cite{ai} and in particular 
\cite{dls} as well as the  contributions to these proceedings 
 by S. de Alwis, J. Liu and
K. Stelle.  In addition, there the  origin of the Randall-Sundrum solution
\cite{rs1} based on $S^1/Z_2$ topology and  two branes  sources was  
addressed in detail.

The outline of this contribution is the following: (i)
We shall  first review the properties of supergravity domain walls,
identifying the  flat (Minkowski) walls with BPS saturated ones
and the bent (deSitter and anti-deSitter) ones with  those whose
energy density is either larger or smaller than that of  the
BPS saturated ones. (This part of the review is primarily based on
a much earlier work on domain walls in  four-dimensional N=1
supergravity,
 initiated in  \cite{cgr}, its generalizations 
to  non-BPS solutions  were given in
 \cite{cgs}  and reviewed in  \cite{cs}; for a recent work on 
a  generalization of  the analysis to D-dimensions see \cite{cw}
and references
 therein.) (ii) We shall then turn to  the  discussion of  the
properties of
domain walls in N=2 D=5 gauged supergravity  (with vector
multiplets), leading to the ``no-go'' theorem for the
implementation of the Randall-Sundrum scenario.  (iii) As the last topic
we turn to the status of the breathing mode domain walls,  where the breathing
mode parameterizes the volume
of the compactified Einstein space (with  the sphere as the most symmetric
example). While these walls may provide  the candidate  gravity trapping
solutions,  we also  mention difficulties with the 
interpreting  of the delta function
sources for infinitely thin domain wall within  Type IIB
string theory.

\section{Domain walls in N=1 supergravity}

Since the scalar potential in supergravity theories is of  the
restricted
form  it is  natural to expect that the nature of domain walls is of
special form as well. The bosonic  Lagrangian  of D=4 N=1
supergravity theory takes the form:
\begin{equation}
{\cal L} = g_{A{\bar B}}\partial_\mu \Phi^A\partial^\mu \Phi^{\bar
B} -V -{R\over {8\pi M_{pl}^2}}\; ,
 \label{lag}
\end{equation}
 where the scalar potential (for the gauge neutral fields)  is of the
form:
 \begin{equation}
 V= g^{A{\bar B}}\partial_A{\hat W} \partial_{\bar B}
  {\hat W} -{3\over M_{pl}^2} {\hat W}^2\, .
 \end{equation}
 Here the K\"ahler invariant quantity ${\hat W}$ 
takes the form:
 ${\hat W}= \zeta |W| e^{K\over {2M_{pl}^2}}$,
 where   $\zeta= \pm  1$ and $\zeta$ can  change the sign  only when   $W$ goes
 through the zero value.  Here
  $W$ and $K$ are the superpotential and  K\"ahler potential, respectively.
Supersymmetric extrema are at $\partial_A {\hat W} =0$,  
and correspond to either a zero cosmological constant,  i.e.
$\Lambda\equiv V_{ext} =0$ 
when ${\hat W}_{ext} =0$,  or a  negative cosmological constant, i.e.
$\Lambda\equiv V_{ext}<0$, when ${\hat W}_{ext}\ne 0$.

The surprising result \cite{cgr}  in the study of domain walls in
supergravity theory
 is that  the static (flat) domain wall  solutions   between supersymmetric
extrema do  exist irrespective of the fact that the actual value of
the cosmological constants of the two isolated extrema  may be different.
 Such solutions turn out to be  BPS ones: they 
satisfy the Killing spinor equations: $\delta
\psi^\alpha =0$,
and $\delta\psi^\alpha_\mu=0$  for the supersymmetric variation of the 
scalar-field super-partner and gravitino, respectively. These equations
are solved with the  static Ansatz for
the scalar field $\Phi$ (only one chosen for simplicity) and the following
conformally flat metric
Ansatz\footnote{We shall describe the  domain wall metric 
interchangeably in
terms of the conformal factor $A$ or the warp factor ${\cal A}$.}:
\begin{equation}
ds^2= A(z) (-dt^2+dx^2+dy^2+dz^2)\equiv {\cal A}({\tilde
z})(-dt^2+dx^2+dy^2)+d{\tilde z}^2
\, , \label{metric}
\end{equation}
leading to the following coupled first order equations of motion \cite{cgr}:
\begin{eqnarray}
\partial_z\Phi^A&=&2 g^{A{\bar B}} \partial_{\bar B}{\hat
W}\, ,\cr
M_{pl}^2\partial_z\log {\cal A}& =&-2 {\hat W}\, ,\label{kink}
\end{eqnarray}
and the energy density of the wall takes the following
  form:
\begin{equation}
\sigma_{BPS}= 2({\hat W}_+-{\hat W}_-)
=2M_{pl}(\zeta_+\sqrt{-{\Lambda_+\over 3}}-\zeta_-\sqrt{-{\lambda_-\over
3}})\, .
\end{equation}

The asymptotic behavior of the solution  is the following:
\begin{eqnarray}
z\to\pm \infty:  \ \ 
 \Phi&\to& \Phi_{\pm}
+e^{2\partial^2_\Phi {\hat W}_{\pm} z}; \ \ \{\partial^2_\Phi
{\hat W}_->0,\partial^2_\Phi {\hat W}_+<0\},\cr 
 {\cal A}
&\to& e^{-{{2{\hat W}_{\pm}z}\over  M_{pl}^2}}. \label{asymp}
\end{eqnarray}
Note that ${\rm sign}({\hat W}_{\pm})$ determines the asymptotic behavior
of
the metric. It turns out that the necessary condition  for the
exponential fall-off  of the  metric  warp factor $\cal A$ is
that at the supersymmetric extrema the potential
  $V(\Phi_{\pm})$  has the minimum.

A typical $Z_2$ symmetric example corresponds to the  following choice:
\begin{equation}
W=\sqrt{\lambda} \Phi({\Phi^2\over 2}-\eta^2),\ \ K=\Phi \Phi^*\, ,
\end{equation}
where the kink solution (\ref{kink}) interpolates between two
supersymmetric anti-deSitter minima with the same cosmological
constant (${\hat W}_+=-{\hat W}_-\equiv {\hat
W}_0=M_{pl}\sqrt{-{\Lambda\over 3}}$). The metric conformal factor
$A(z)$  (\ref{metric}) falls-off on either side  as $M_{pl}^4({\hat W}_0
z)^{-2}$, and the
energy density of the wall is  $\sigma_{BPS} =2 \times 2
{\hat W}_0=4M_{pl}\sqrt{-{\Lambda\over 3}}$.

The thin wall  limit of the  $Z_2$ symmetric  BPS solution is achieved
by taking the limit
$\lambda\to \infty$, $\eta\to 0$, while $\lambda\eta^3$ remains
fixed. In this particular case the superpotential $\hat W$
approaches the step function, while the potential and the scalar
kinetic energy term approach the delta
function. This 
leads to the thin wall effective (supersymmetric) Lagrangian
with the delta function source that precisely reproduces the
thin- wall flat domain-wall solution, discussed by
Randall-Sundrum in
D=5.  (For further discussions for  a  supersymmetric implementation 
of the  effective Lagrangian with  the delta function sources
 see \cite{gp,abn,flp,bkp,cy}.)

In the case of broken supersymmetry, the walls become bent. In
the thin wall analysis of  walls  that
can assume the $Z_2$ symmetric limit,  the  space-time internal
to the wall is either deSitter (when $\sigma > \sigma_{BPS}$) or
anti-deSitter (when $\sigma <\sigma_{BPS}$).  The conformal
factor (\label(\ref{metric}) on the other hand takes the form \cite{cgs}:
 \begin{eqnarray} A(z)&=&
M_{pl}^2\beta^2\left[\sqrt{-{\Lambda_{\pm}\over 3}}
\sinh(\beta(z+z_{\pm}))\right]^{-2}, \ {\rm  for}\ \ 
\sigma>\sigma_{BPS}\, , \cr
A(z)&=&M_{pl}^2\beta^2\left[\sqrt{-{\Lambda_{\pm}\over 3}}
\cos(\beta(z+z_{\pm}))\right]^{-2}, \ {\rm for} \ \ 
\sigma<\sigma_{BPS}. \label{bentmetric}
\end{eqnarray}
where $\beta^2\sim |\Lambda_{wall}|$  \
specifies the cosmological
constant on the bent  wall for respective anti-deSitter and
deSitter space-times on the  wall.  The constants $z_{\pm}$ are
appropriately chosen so that at $z=0$ the conformal factor is
$A(z)$ is normalized to 1. In the former case the conformal
factor falls-off even faster then in the  BPS case ($|z|\to
\infty$ corresponds to
the cosmological  horizons), while for the
latter case the conformal factor, while first decreasing it turns
around at intermediate distances ($|z|\to \infty$ is the  space-time boundary).
This latter case allows for the 
possibility of quasi-localized gravity on the wall \cite{kr}.

While we have briefly reviewed the  BPS and non-BPS thin walls 
for supergravity theory in  four dimensions, it was
shown  in \cite{cw} that the global and local space-time structure of
co-dimension one objects in D dimensions is universal, only that
the
role of the cosmological constant factor
$\sqrt{-{{\Lambda}_\pm\over 3}}$ is replaced by
$\sqrt{-{(D-2){\Lambda}_\pm\over{2(D-1)}}}$. In particular, for
domain walls in D=5, this factor becomes
$\sqrt{-{3{\Lambda}_\pm\over{8}}}$.

\section{Domain Walls in D=5 Gauged Supergravity}

A natural question to be asked is to identify the
origin of gravity trapping solutions in fundamental (M-)theory,  as initiated in
\cite{v}.
In particular, can such domain walls arise in an effective 
five-dimensional theory, that can be obtained as a compactification
of, e.g.,  M-/string theory on Einstein-Sasaki spaces?
 Since such 
compactifications are expected to  produce an effective gauged supergravity
theory, we
now turn to the  study of domain walls  in five-dimensional N=2 gauged
supergravity.  For the sake of simplicity, we shall focus on D=5 N=2
gauged supergravity  with Abelian ($U(1)_R$)-gauging and with the
vector supermultiplets, only \cite{gst}.

The bosonic sector is of the constrained form, with the vector
supermultiplets $X^I$ (real, neutral fields) subject to the
 following condition:
\begin{equation}
F=\sum_{IKJ} C_{IJK} X^I X^J X^K=1 \, ,
\end{equation}
which can be
solved for the physical scalar fields $\Phi^A$. The  potential of $U(1)$ gauged
supergravity is also of the constrained form:
\begin{equation}
W=\sum_I h_I X^I \, .
\end{equation}
The bosonic part of the Lagrangian nevertheless takes an
analogous form as in the case of D=4 N=1 supergravity (We have
now  set $M_{pl}$=1.):
\begin{equation}
{\cal L}= g_{AB}\partial_\mu  \Phi^A\partial^\mu \Phi^B -V +R\, ,
\end{equation}
where the metric and the potential are of the form:
\begin{equation}
g_{AB}= {1\over 2} (\partial_I\partial_J F)\partial_A
X^I\partial_B X^J; \ \ V=g^{AB} \partial_AW\partial_BW-{4\over 3}
W^2\, . \end{equation}
Again, the Killing spinor equations, corresponding
to the supersymmetric domain wall solutions,  reduce to the
equations of the analogous type:
\begin{eqnarray}
\partial_z\Phi^A&= &\mp 3g^{AB} \partial_B W\, , \cr
\partial_z\log {\cal A} &=&\pm 2 W \, . 
\end{eqnarray}
 However, due to the fact that $W$ and $g_{AB}$ are of the
constrained form, they 
satisfy a  relationship \cite{gst}, that
takes the  following  form at the supersymmetric extrema:
\begin{equation}
\partial_A\partial_B W_{ext} =\textstyle{2\over 3} [g_{AB}W]_{ext}\,
.\label{susyrel}
\end{equation}
Expanding  the Killing spinor equations around the supersymmetric
minima $\partial_{\Phi}W|_{\pm}=0$, as  $\Phi^A=
\Phi^A_{\pm}+\delta \Phi^A$ and using the relationship
(\ref{susyrel}), yields the following  asymptotic form of these
equations:
\begin{eqnarray}
\partial_z(\delta \Phi^A)& =&\mp W_{\pm}\delta\Phi^A \, ,\cr
\partial_z(\log {\cal A}) &=&\pm W_\pm \, .\label{symeq}
\end{eqnarray}
The condition for the asymptotic kink solution requires
${\rm sign}( W_+)=-{\rm sign} (W_-)$, and then  the
conspiracy of signs
in the above equations (\ref{symeq})  necessarily requires that  the kink
solution has the metric factor that 
{\it grows} exponentially on either side of the
wall. Thus, these solutions 
  are not relevant for the
trapping of gravity; those are typical domain walls  relevant  for 
 AdS/CFT correspondence.  In addition, the constraint
(\ref{susyrel}) implies that   supersymmetric extrema with $W_{ext}>0$ are
necessarily  the minima of the superpotential
 (for $g_{AB}$-positive definite), while  
 supersymmetric extrema with $W_{ext}< 0$ are maxima. 
 Therefore, the kink necessarily
has to cross the singular region in the superpotential manifold
and thus the  solution is generically singular.

Further studies  of more general solutions reveal that neither
inclusion of non-Abelian tensor multiplets \cite{gz,kl}, nor
inclusion of hypermultiplets \cite{cd,bhl}  allow for non-singular  domain
walls  that would have a
 fall-off metric conformal factor on both sides of the
wall. (For a related no-go theorem see \cite{mn}.)

\section{Breathing mode and gravity trapping domain walls}

In this section we shall  review 
a framework
within gauged supergravity theories that has a chance of
implementing  a variant of the Randall-Sundrum scenario. (For related work see,
e.g., 
\cite{cpv}.)  Recall that
in order to obtain  $Z_2$ symmetric  domain-wall solutions of the
Randall-Sundrum
scenario, the gauged supergravity potential would have to have
two isolated supersymmetric {\it minima}. Since the potentials
for the massless scalar fields in a gauged supergravity  generically do not
have this feature, we  now turn to an alternative proposal \cite{clp1}
 to include other scalar fields that do not lie in the massless
supermultiplet.

We shall focus on the special classes of gauged supergravities that
arise from sphere reductions of M-theory or string theory, with
particular emphasis on the $D=5$ case.  For examples in the
Kaluza-Klein reduction of Type IIB string theory on a five-sphere
($S^5$), there will be an infinite tower of massive supermultiplets in
addition to the massless multiplet, and so one could consider the
potentials for one or more of the massive scalar fields.  In general,
one cannot focus attention on a single such field in isolation, on
account of its couplings to other fields.  However, in certain special
cases a consistent truncation to a single massive scalar can be
performed.  One such example is the ``breathing mode'' that
parameterises the overall volume of the compactifying $S^5$.  (Unlike
the breathing mode in a toroidal reduction, which is massless, the
breathing mode in a spherical reduction is a member of a massive
supermultiplet.)

   The scalar potentials for the breathing-mode scalars in various
Kaluza-Klein spherical reductions were studied in \cite{bdlps}.
Although the breathing mode is a member of a massive multiplet, the
truncation is nonetheless consistent since it is a singlet under the
isometry group of the internal sphere. (It would not in general be
consistent to turn on a finite subset of other fields as well.)

The resulting $D$-dimensional
Lagrangians all turn out to have the following form:
\be
{\cal L}_D = \, R - \ft12 \, (\del\phi)^2 - \, V\,,
\ee
where the potential is given by \cite{bdlps}
\be V=\ft12\, g^2(\fft{1}{a_1^2}\, e^{a_1\Phi} - \fft{1}{a_1
a_2}\, e^{a_2\Phi})\,.\label{scalarpot} \ee
The positive constants $a_1$ and $a_2$ are given by
\be a_1^2 = \fft{4}{N} + \fft{2(D-1)}{D-2}\,,\qquad a_1\, a_2=
\fft{2(D-1)}{D-2}\,, \ee
where Type IIB reduction on $S^5$ corresponds to $N=1$, $D=5$.
 Since  $a_1>a_2>0$, the
potential has a minimum at $\phi=0$, corresponding to the
self-dual point  where the volume of the five-sphere and the
radius of the AdS$_5$ are equal (in appropriate units). In
addition, this potential can be cast in the  standard
supersymmetric form:
\be
V=(\fft{\del W}{\del \phi})^2 -\fft{D-1}{2(D-2)}\, W^2\,,
\ee
where
\be W=\sqrt{\fft{N}{2}}\, g\, (\fft{1}{a_1}\, e^{a_1\Phi/2}
- \fft1{a_2}\, e^{a_2\Phi/2})\,. \ee
Thus,  there is  a domain wall solution that can be obtained 
 in terms of the coupled
first-order differential equations (\ref{symeq}).
Solving for $\phi$ and $A$ one finds the result:
\be A^{{D-1}\over 2} =  c\, \fft{\del W}{\del \Phi}\, e^{-\fft12
(a_1+a_2)\Phi}\,,\label{afromphi} \ee
where $c$ is an integration constant and the solution for $\Phi$
is given by:
\be z-z_0 = \fft4{a_2\, g\, \sqrt N}\, e^{-\fft12 a_2\, \Phi}\,
{}_2F_1[\fft{a_2}{a_2-a_1}, 1, 1+\fft{a_2}{a_2-a_1}; e^{\fft12
(a_1-a_2)\, \Phi}]\,.\label{hype} \ee

    The  solutions above  {\it have two different branches}. In one
branch, $\phi$ runs from 0 to $+\infty$, with $z$ running from
$z=-\infty$ to $z=0$, where we have chosen the integration
constant $z_0$  to yield the following result:
\bea
e^{-\fft12 a_1\, \phi} &\sim& -\ft14 a_1\,\sqrt{N}\, g\, z\,,\nn\\
A^{{(D-1)}\over 2} &\sim& c\, \sqrt{\fft{N}{8}}\, g\, e^{-\fft12 a_2\,
\phi} \sim  c\, \sqrt{\fft{N}{8}}\, g\,\Big(- \ft14 a_1\, \sqrt{N}\,
g\, z\Big)^{\fft{a_2}{a_1}}\,.
\eea
In this branch, when the coordinate $z$ reaches its limit at $z=0$,
the  metric factor  therefore goes to zero, and there is a
power-law naked curvature singularity.
(Note that in this regime the solution  extends
into  large positive values of the  potential (\ref{scalarpot}) with a
large cost to the energy density of the wall, and it
thus terminates at a finite value of the transverse coordinate.)
As $z$ approaches $-\infty$, the  metric  asymptotically approaches the  AdS  space-time, described in
horospherical coordinates with $z\to -\infty$ corresponding to the Cauchy
horizon. Note that on that side of the wall the gravity is
repulsive and  provides ``one half'' of the Randall-Sundrum wall.

The study of the gravitational fluctuating modes, 
internal to
the wall, reduces to the  study of  the Schr\" odinger equation, whose 
 potential  \cite{clp2} has an attractive, singular region 
 near the naked singularity. Nevertheless the  spectrum  has energy levels 
 bounded from below.  
However, the boundary conditions  at the naked singularity 
exclude the massless normalizable mode 
(corresponding to  the four-dimensional
graviton).  On the other hand corrections  (of the order of the
inverse string
scale), that would  smooth out the naked singularity, would
in turn provide the  non-singular attractive  Schr\"odinger
potential   with precisely one normalizable massless state.
Further investigations  to identify the  origin of the smoothing  out 
of  such singularities within
the string theory context is needed.

In the second branch, $\phi$ runs from 0 to $-\infty$, while $z$
runs from $z=-\infty$ to $z=+\infty$.  The behaviour of the solution
near $z=-\infty$ is the same as in the branch discussed previously,
with the metric approaching  asymptotically AdS.
 As $z$ approaches $+\infty$, the
solution becomes
\bea
e^{-\fft12 a_2\, \phi} &\sim& \ft14 a_2\,\sqrt{N}\, g\, z\,,\nn\\
A^{{(D-1)}\over 2} &\sim& - c\, \sqrt{\fft{N}{8}}\, g\, e^{-\fft12 a_1\,
\phi} \sim  - c\, \sqrt{\fft{N}{8}}\, g\,\Big( \ft14 a_2\, \sqrt{N}\,
g\, z\Big)^{\fft{a_1}{a_2}}\,.
\eea
(The constant $c$ is negative in this case.)  This side describes
one-side of a supersymmetric {\it dilatonic} domain wall.
Interestingly, it has {no curvature singularity}; as $z$ tends to
$+\infty$ the curvature falls off as $1/z^2$, while the diverging
dilaton $\phi\to -\infty$ approaches the weak coupling limit.  
Unfortunately, the dilatonic vacuum side does not provide for  the  gravity trapping
solution.

Thus within a pure field-theoretic framework, i.e. by  employing
only the
breathing-mode scalar field to construct the 
domain wall solution,  one did not
 fully succeed in constructing domain wall solutions  that would allow for trapping of
gravity on the wall, though the first branch  would provide 
for  such a
scenario if   the mechanism to smooth out 
 the naked singularity  within string
theory existed. (For a possible related mechanism see
\cite{jpp}.) 

In \cite{clp1} it  was therefore proposed  to add a
delta-function source  to  the effective Lagrangian. In this case, 
  the diverging
behaviour of
the dilaton is  cut-off  by a delta-function source 
at some finite value of $z$, say $z= z_*$ and  the solution for $z > z_*$ can be
replaced by a reflection of the solution for $z< z_*$. 
Now the solution is $Z_2$ symmetric and  the metric
factor $A(z)$  falls-off on either  side of the infinitely thin
wall, supported
by the delta function source, thus reproducing
the Randall-Sundrum scenario.
The origin of such a  delta function source  within Type IIB
supergravity  was further explored in  \cite{cdllps}.  

\vskip 0.5cm

{\bf Acknowledgments} I would like to thank K. Behrndt, H. L\" u and C. Pope for many
discussions and enjoyable collaboration on topics presented in this
contribution. The work is supported by U.S. Department of Energy Grant No.
DOE-EY-76-02-3071 and by the NATO Linkage grant 97061.

\end{document}